# LENSE THIRRING AND GEODETIC EFFECTS

M. Cattani
Instituto de Fisica, Universidade de S. Paulo, C.P. 66318, CEP 05315−970
S. Paulo, S.P. Brazil . E−mail: mcattani@if.usp.br

Abstract.
Using the Einstein gravitation theory (EGT), we analyze the Lense Thirring (LT) and the Geodetic effects. In the LT effect the angular orbital momentum **L** and the perigeo of a particle, orbiting a sphere with mass M and spin **J** = I$\omega$ around an axis passing by its center of mass, precess around **J**. In the Geodetic effect the spin **S** of a gyroscope orbiting M precess around its orbital angular momentum **L** and the spin of **J** of M. The theoretical predictions are compared with the experimental results. This article was written to graduate and postgraduate students of Physics.

Resumo.
Usando a teoria de gravitação de Einstein (TGE) estudamos os efeitos Lense Thirring (LT) e Geodético. No efeito LT o momento angular **L** e o perigeo de uma partícula, orbitando uma esfera com massa M e spin **J** = I$\omega$ em torno de um eixo que passa pelo seu centro de massa, precessionam em torno de **J**. No efeito Geodético o spin **S** de um giroscópio orbitando M precessiona em torno de seu próprio momento angular orbital **L** e do spin **J** de M. As previsões teóricas são comparadas com resultados experimentais. Esse artigo foi escrito para alunos de graduação e pós−graduação de Física.

## I) Introdução

Num artigo precedente[1a] usando a TGE calculamos a métrica do espaço−tempo denominada de métrica de Schwarzschild (MS) que é gerada no vácuo ao redor de uma distribuição esfericamente simétrica de massa M, sem carga e não em rotação. A massa M está em repouso na origem O de um referencial inercial. A MS em coordenadas polares (r,$\theta$,$\varphi$) é definida através do invariante ds$^2$ dado por

$$ds^2 = (1 - 2GM/c^2r) \, c^2dt^2 - dr^2/(1 - 2GM/c^2r) - r^2d\theta^2 - r^2\sin^2\theta \, d\varphi^2 \quad (I.1),$$

que no limite de campos fracos é dado, em coordenadas cartesianas (X,Y,Z), por

$$ds^2 \approx (1 - 2GM/c^2r) \, c^2dt^2 - (1 + 2GM/c^2r)(dX^2 + dY^2 + dZ^2) \quad (I.2),$$

onde $r = (X^2+Y^2+Z^2)^{1/2}$. Usando a MS testamos[1a–1c] previsões feitas pela TGE analisando vários fenômenos [2–5] tais como dilação temporal, deflexão e efeito Doppler da luz, precessão do periélio de planetas, desvios da teoria Newtoniana nos movimentos planetários, atraso temporal de ecos de sinais de radar passando ao redor do Sol, lentes gravitacionais e emissão de ondas gravitacionais pelo binário constituído pelo pulsar PSR 1913+ 16.

Veremos agora o *Efeito Lense−Thirring* (L−T)[6] e o *Efeito Geodético* (EG) ou de *de Sitter*[2,3]. No efeito L−T o momento angular orbital **L** e do perigeo de uma partícula orbitando uma esfera com massa M, que gira com velocidade angular ω em torno de um eixo que passa pelo seu centro de massa O, precessionam em torno do momento angular do ("spin") **J** = Iω de M. No *Efeito Geodético* (EG) ou de *de Sitter*[2,3] o spin **S** de um giroscópio orbitando M precessiona em torno de seu próprio momento angular orbital **L** e de **J**. A massa M está em repouso em um referencial inercial com coordenadas cartesianas (X,Y,Z) com os respectivos versores (*x,y,z*) com origem O no centro da esfera. Seja **J** = Iω = Iω*z* o momento angular de M ao longo do eixo Z e $I = (2/5)MR^2$ o seu momento de inércia. A solução exata das equações de campo da TGE para o caso da massa em rotação foi obtida por Kerr.[6–8] A expressão exata para o intervalo $ds^2$ pode ser vista, por exemplo, no Ohanian[3] (pág.324). A métrica correspondente que define a geometria de Kerr é conhecida como *Métrica de Kerr* (MK). Ela é uma solução estacionária da TGE, mas não estática, isto é, a MK não é uma função de t, mas não é invariante por uma reflexão temporal devido a presença de termos mistos do tipo $dtdX^i$. Ela é simétrica por uma rotação em torno do eixo Z, isto é, é invariante por uma transformação φ → φ + dφ. A geometria de Kerr que é muito mais complicada do que a geometria de Schwarzschild[3] é de importância crucial para se analisar buracos negros sem carga e em rotação.

No caso limite de valores r grandes, que nos interessa, o invariante $ds^2$ na MK é dado por [3,10]

$$ds^2 \approx (1-2GM/c^2r)\,c^2dt^2 - (1+2GM/c^2r)\,dr^2 - r^2d\theta^2 - r^2\sin^2\theta\,d\varphi^2 - (4GI\omega/c^2r)\sin^2\theta\,d\varphi\,dt$$

ou, em coordenadas cartesianas (X,Y,Z),[5,11]  (I.2)

$$ds^2 \approx (1-2GM/c^2r)\,c^2dt^2 - (1+2GM/c^2r)(dX^2 + dY^2 + dZ^2) + g_{14}\,dXdt + g_{24}\,dYdt,$$

levando em conta que $r^2\sin^2\theta\,d\varphi = XdY - YdX$, onde $g_{41} = g_{14} = g_{42} = g_{24} = +2GI\omega/c^2r^2$ e $r = (X^2+Y^2+Z^2)^{1/2}$. A (I.2) é a *métrica linearizada* de Kerr. Ela descreve o campo gravitacional no vácuo a grandes distâncias de uma estrela ou planeta, em rotação e sem carga.

# 1) Efeito Lense–Thirring.

Consideremos agora uma partícula com massa m (satélite) orbitando a esfera de massa M e raio R, que suporemos ser a Terra (em repouso em O), que está em rotação em torno do eixo (norte-sul Z), conforme Figura 1. Segundo artigos anteriores,[1b,1c] se a esfera não girasse a partícula m descreveria uma órbita elíptica contida em um plano fixo no espaço. Veremos agora o que ocorre com essa trajetória quando a esfera está em rotação em torno de um eixo Z com momento angular orbital **J** = I$\omega$ (vide Fig.1), onde **ω** é a velocidade angular de rotação da esfera em torno de Z. O movimento de m será observado do sistema inercial com origem em O. A órbita de m dada por **r** = **r**(t) será descrita em coordenadas polares esféricas (r,θ,φ) e tomando como base o referencial inercial (X,Y,Z) conforme Figura 1. Os versores polares serão indicados por ***r***, ***θ*** e ***φ***, respectivamente.

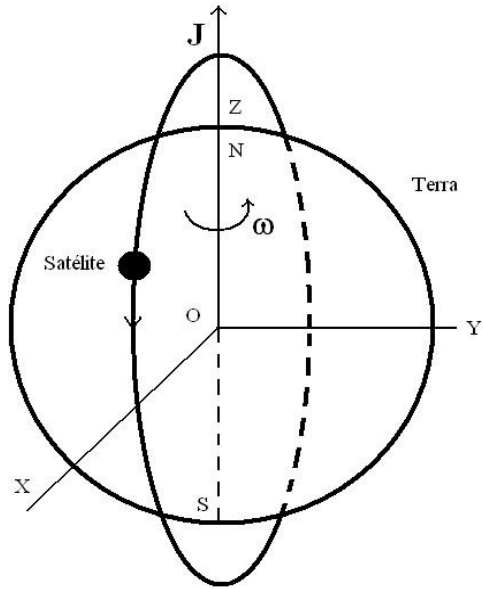

**Figura 1**. Mostramos um satélite orbitando a Terra. O centro de massa da Terra está em O e ela gira em torno do eixo Z (norte –sul) com momento angular ("spin") **J** = I **ω**.

Assumiremos que os efeitos relativísticos sejam pequenos[1–5,10,11] (velocidades baixas e campos gravitacionais fracos). Neste caso a métrica $g_{\mu\nu}$ do espaço-tempo diferindo muito pouco da métrica de Minkowski $g_{\mu\nu}^{(o)}$ = (–1,–1,–1, 1) será dada por $g_{\mu\nu} = g_{\mu\nu}^{(o)} + h_{\mu\nu}$ onde $h_{\mu\nu} = h_{\nu\mu}$ é uma pequena perturbação de $g_{\mu\nu}^{(o)}$. Nessas condições os símbolos[1] de Christoffel $\Gamma_{\mu\nu}^{\alpha}$ ficam escritos como[1–5,10]

$$\Gamma_{\mu\nu}^{\rho} = (g^{\rho\lambda}/2)(\partial_\nu g_{\lambda\mu} + \partial_\mu g_{\lambda\nu} - \partial_\lambda g_{\mu\nu}) = (h^{\rho}{}_{\mu,\nu} + h^{\rho}{}_{\nu,\mu} - h_{\mu\nu}{}^{,\rho}) \qquad (1.1).$$

A geodésica de m em torno de M é dada por[1d]

$$d^2x^\rho/ds^2 + \Gamma_{\mu\nu}{}^\rho (dx^\mu/ds)(dx^\nu/ds) = 0 \qquad (1.2).$$

Assim, definindo $\gamma = dt/ds = c\,[1-(v/c)^2]^{-1/2}$, teremos

$dx^\rho/ds = (dx^\rho/dt)(dt/ds) = (dx^\rho/dt)\,\gamma$

e

$d^2x^\rho/ds^2 = \gamma^2 (d^2x^\rho/dt^2) + (dx^\rho/dt)(d\gamma/dt)\gamma$ .

Assumindo que a velocidade v do satélite seja v << c, desprezando termos em segunda ordem $(v/c)^2$ temos $\gamma^2 \approx c^2$ e $(d\gamma/dt)\gamma \approx 0$. Como nessas condições $d^2x^\rho/ds^2 \approx (d^2x^\rho/dt^2)c^2$ a (1.2), usando (1.1), fica[5,10]

$$d^2x^\rho/dt^2 = -\Gamma_{\mu\nu}{}^\rho (dx^\mu/dt)(dx^\nu/dt) = -(h^\rho{}_{\mu,\nu} + h^\rho{}_{\nu,\mu} - h_{\mu\nu}{}^{,\rho})(dx^\mu/dt)(dx^\nu/dt) \quad (1.3).$$

Conforme (I.2), $h^{44} = -2GM/c^2r$, $h_{11}=2GM/c^2r$, $h_{41} = h_{14} = h_{42} = h_{24} = 2GI\omega/c^2r^2$, onde $r = (X^2+Y^2+Z^2)^{1/2}$. Os demais $h_{\mu\nu}$ são nulos. Como todos os $h_{\mu\nu}$ são independentes do tempo, ou seja, $h_{\mu\nu,4} = 0$ podemos mostrar, usando (1.3) que[10]

$$\mathbf{a} = d^2\mathbf{x}/dt^2 = d\mathbf{v}/dt = -\operatorname{grad}(\Phi) + \mathbf{v} \times \operatorname{rot}(\mathbf{W}) \qquad (1.4),$$

onde

$\Phi = -GM/r$ é o potencial Newtoniano

e

$\mathbf{W} = -4G(\mathbf{J} \times \mathbf{r})/c^2r^3 = -4G(I\boldsymbol{\omega} \times \mathbf{r})/c^2r^3$.

De acordo com a Mecânica Clássica,[12,13] quando um corpo de massa m se movimenta com velocidade **v** em relação a um referencial não−inercial que gira com velocidade angular **Ω** em torno de um eixo verificamos que sobre o corpo aparece uma força adicional não−inercial dada por 2 m **v** x **Ω** denominada de "força de Coriolis". Assim, se sobre m há uma força aplicada $-m\,\operatorname{grad}(\Phi)$ a força resultante **f** sobre um corpo será dada por

$$\mathbf{f} = m\mathbf{a} = -m\,\operatorname{grad}(\Phi) + 2m\,\mathbf{v} \times \boldsymbol{\Omega} \qquad (1.5).$$

De (1.4) e (1.5) pode−se dizer que em um campo gravitacional criado por um corpo em rotação (com momento angular **J**) uma partícula muito distante do corpo está submetida a uma força equivalente à força de Coriolis devido a uma rotação com velocidade angular **Ω** dada por

$$\boldsymbol{\Omega} = (1/2)\operatorname{rot}(\mathbf{W}) = (G/c^2r^3)\,[\mathbf{J} - 3\mathbf{r}(\mathbf{J}\cdot\mathbf{r})] \qquad (1.6),$$

onde *r* = **r**/r é o versor ao longo da direção radial. Nesse ponto consideraremos o seguinte resultado da Mecânica Clássica. Seja **G** um vetor arbitrário definido num sistema inercial Σ e (d**G**/dt)$_Σ$ a sua variação temporal em Σ. A sua variação temporal (d**G**/dt)$_{rot}$ em relação a um sistema não−inercial que gira com velocidade angular **Ω** em torno de um eixo é dada por[12,13]

$$(d\mathbf{G}/dt)_{rot} = (d\mathbf{G}/dt)_Σ + \mathbf{Ω} \times \mathbf{G} \quad\quad (1.7).$$

Como **G** é arbitrário a (1.7) define de fato uma transformação da derivada temporal d/dt entre dois sistemas de coordenadas, um inercial e outro em rotação não−inercial. No caso particular em que **G** é constante em Σ, ou seja, (d**G**/dt)$_Σ$ = 0 teremos simplesmente

$$(d\mathbf{G}/dt)_{rot} = \mathbf{Ω} \times \mathbf{G} \quad\quad (1.8).$$

Tendo em vista (1.1)−(1.8) vemos que o spin **J** da Terra causa uma precessão do momento angular **L** do satélite dada por

$$d\mathbf{L}/dt = \mathbf{Ω}_{LT} \times \mathbf{L} \quad , \quad\quad (1.9),$$

onde a velocidade angular **Ω**$_{LT}$ de precessão é dada por

$$\mathbf{Ω}_{LT} = (G/c^2 r^3)\,[\mathbf{J} - 3\mathit{r}(\mathbf{J}\cdot\mathit{r})] \times \mathbf{L} \quad\quad (1.10),$$

e é denominada *velocidade angular de precessão de Lense−Thirring*. O plano orbital da particular teste seria um enorme giroscópio submetido à ação dos efeitos do spin **J** da massa M. A órbita da partícula teste em volta do spin **J** teria uma mudança secular de longitude da linha de nós (interseção entre o plano da órbita e o plano equatorial da massa M). É como se a rotação da Terra arrastasse o plano orbital do satélite. Por essa razão o efeito L−T é também denominado de "orbital dragging effect" ou "dragging inertial frame effect".

Analisemos agora a precessão **Ω**$_{LT}$ dada por (1.10), que escreveremos na forma **Ω**$_{LT}$ = (GJ/c²r³) [*z* − 3*r*(*z*·*r*)] × **L**, em dois casos particulares extremos.

(1) *Trajetória da partícula no plano equatorial perpendicular à **J**.*
Nesse caso como *z* × **L** = 0 e *z*·*r* = 0 a (1.11) fica dada por

$$d\mathbf{L}/dt = 0,$$

ou seja, **Ω**$_{LT}$ = 0 e não há precessão de **L**.

(2) *Trajetória **r**(t) da partícula e **J** num mesmo plano equatorial.*

Nesse caso como **z** x **L** = ± L**φ**, onde **φ** é o versor ao longo dos ângulos φ, perpendicular à trajetória, **z·r** = cosθ(t) e **r** x **L** = ± L **θ** a (1.11) fica dada por

$$d\mathbf{L}/dt = (GJ/c^2r^3)[\pm\boldsymbol{\varphi} - 3\cos\theta(t)\boldsymbol{\theta}]L \qquad (1.11),$$

mostrando que há uma precessão de **L** em torno de **J**. O sinal ± da precessão depende do sentido horário ou anti−horário de **L**. Como estamos preocupados com o efeito secular da precessão, levando em conta que valor médio no tempo <cosθ(t)> = 0 e (1.11) temos d**L**/dt = ±(G/c²r³)JL**φ** = **Ω**$_{LT}$ x **L**. Obtendo assim a velocidade angular de precessão **Ω**$_{LT}$ de **L** em torno de J, dada por

$$\boldsymbol{\Omega}_{LT} = (G/c^2r^3)\mathbf{J} \qquad (1.12).$$

Levando em conta ainda que a órbita de m é uma elipse com semi−eixo maior a e ecentricidade ε e calculando a média temporal[5,13,14] de 1/r³(t), obtemos

$$\Omega_{LT} = GJ/[c^2a^3(1-\varepsilon^2)^{3/2}] \qquad (1.13),$$

onde $J = S_E = I\omega = (2/5)MR^2\omega$.

## 1.a) Precessão do Perigeo da Órbita da Partícula devido ao Spin J.

Além da precessão de **L**, vista na Seção 1, a interação gravitacional causa uma precessão do perigeo da órbita (plana) da partícula descrita em torno da massa M. É oportuno lembrar que no referencial inercial definido sobre o plano da órbita temos duas grandezas invariantes[5,12,13] que são o momento angular **L** e o vetor axial **A** ("Vetor de Lagrange−Runge−Lenz") dados por

$$\mathbf{L} = m\mathbf{v} \times \mathbf{r} \quad \text{e} \quad \mathbf{A} = (\mathbf{p}/m) \times \mathbf{L} - 2GMm\mathbf{r}/r. \qquad (1.14).$$

O vetor **L** é perpendicular ao plano da órbita e **A** é dirigido ao longo do semi−eixo maior da elípse no sentido do perigeo[5,13] (|**A**|=εGMm).

Ora, de (1.12) podemos concluir[5,12,13] que a precessão de **L** em torno de **J** é devida a uma interação U = 2G**L J**/c²r³. Além da precessão essa interação gera um efeito radial que irá causar a precessão do perigeo da órbita de m em torno de M. O aparecimento de uma força radial pode ser constatado verificando, usando (1.5) e (1.6), que a força **f** apresenta uma componente radial além da gravitacional − m grad(Φ). No caso particular, por exemplo, em que a trajetória da partícula está no plano equatorial perpendicular à **J**, ou seja, quando **J·r** = 0, verificamos que

$$\mathbf{f} = m\,[-(GM/r^2) \pm (6GJ/c^2r^3)]\,\mathbf{r}. \qquad (1.15).$$

Nessas condições, devido a força radial $6GJ/c^2r^3$, usando a mecânica clássica, podemos mostrar que o perigeo de uma trajetória elíptica (aproximadamente circular) em torno da massa M teria uma velocidade angular de precessão de $\mathbf{\Omega}_P$ dada por[5,15] $\mathbf{\Omega}_P \approx -(6GJ/c^2a^3)(\mathbf{L}/L)$, onde a $\approx$ semi−eixo maior $\approx$ raio da órbita e $\mathbf{L} \approx ma^2\omega^*$, sendo $\omega^*$ a velocidade angular de m em torno de M. Note que para calcular a precessão do perigeo $\mathbf{\Omega}_P$ não levamos em conta a precessão dada pela TGE como vimos num artigo anterior[1b]. De acordo com a TGE[1b] essa precessão seria dada por $\Omega_{TGE} \sim 3GM/ac^2$ e conforme cálculos acima a precessão de *arrasto* $\Omega_P$ é dada por $\Omega_P \sim 3GJ/\omega^*a^3c^2$ onde $J = I\omega = (2/5)MR^2\omega$. Desse modo a razão $\Omega_{TGE}/\Omega_P \sim 2\omega(R/a)^2/5\omega^*$. Como para os satélites LAGEOS usados[18] para medir o efeito L−T, a $\approx$ 2R e $\omega^* \approx 4\omega$ temos $\Omega_P \sim 2.5\,\Omega_{TGE}$. Pode−se mostrar[5,16,17] de modo mais rigoroso que o satélite descreveria uma elipse cujo perigeo sofreria uma precessão de *arrasto* $\mathbf{\Omega}_P$ dada por

$$\mathbf{\Omega}_P \approx -(3G/c^2r^3)\,(\mathbf{J}\cdot\mathbf{L}/L^2)\mathbf{L}. \qquad (1.16).$$

Sob a ação do arrasto devido a **J** o momento angular **L** da partícula precessionaria em torno de **J** com velocidade angular $\mathbf{\Omega}_L = \mathbf{\Omega}_{LT} = (G\mathbf{J}/c^2r^3)$. Levando em conta ambos os arrastos descritos por $\mathbf{\Omega}_L = \mathbf{\Omega}_{LT}$ e $\mathbf{\Omega}_P$ a precessão do vetor **A** (precessão do perigeo), seria dada por

$$d\mathbf{A}/dt = \mathbf{\Omega}_A \times \mathbf{A}, \qquad (1.17)$$

onde

$$\mathbf{\Omega}_A = \mathbf{\Omega}_L + \mathbf{\Omega}_P = (G/c^2r^3)[\mathbf{J} - 3\mathbf{L}(\mathbf{J}\cdot\mathbf{L}/L^2)].$$

De modo análogo a (1.13) a variação secular de $\mathbf{\Omega}_A$ é dada por

$$\mathbf{\Omega}_A = G[\mathbf{J} - 3\mathbf{L}(\mathbf{J}\cdot\mathbf{L}/L^2)]/(c^2a^3(1-\varepsilon^2)^{3/2}) \qquad (1.18).$$

**1.b) Comparação com Resultados Experimentais. Conclusões.**

A detecção e medidas do efeito L−T foram feitas[18] usando os satélites LAGEOS ("laser geodynamics satellite") da NASA and LAGEOS II (NASA e ASI, Agência Espacial Italiana) adotando os modelos de campo gravitacional da Terra JGM−3 e EGM−96. No caso do satélite LAGEOS lançado em 1976 a velocidade de precessão $\Omega_L$, de acordo com (1.13), seria de $\Omega_L \approx 0.030"/y$ e para o LAGEOS II, lançado em 1992, seria de $\Omega_L \approx 0.031"/yr$. As precessões $\Omega_A$ calculadas com a (1.18) para os satélites LAGEOS e LAGEOS II são $\Omega_A \approx .032"/yr$ e $\Omega_A \approx -.057"/y$, respectivamente. Levando em conta a análise das órbitas feitas pelos satélites LAGEOS e LAGEOS II verificou−se concluir que o efeito L−T existe, dentro de um limite de 10%, e que o valores medidos de Ω concordam com um erro de ± 20−30% com o que é previsto usando a TGE, conforme (1.13) e (1.18). É importante lembrar que o arrasto do plano da órbita do satélite devido ao momento angular intrínseco **J** do corpo central está de acordo com a formulação geral relativística do Princípio de Mach.[2]

### 3) Efeito Geodético.

Pugh[19] e Schiff[20], em 1959−1960, sugeriram que giroscópios *esféricos* colocados em satélites em órbita (geodésica) ao redor da Terra poderiam dar informações muito precisas sobre o campo gravitacional terrestre medindo−se as precessões de seus spins. Essa precessão é conhecida como *Efeito Geodético* ou de *de Sitter*.[21] As coordenadas com origem O no referencial inercial no centro da Terra serão indicadas por $X^\mu$.

Vamos assumir que um giroscópio esférico esteja na origem de um *referencial geodésico* (RG) (vide Apêndice A) ou *referencial em queda livre* (RQL) ou, ainda, *comoving referential system* (CRS). Neste referencial a variação do momento angular (spin), que indicaremos por $S^\mu$, do giroscópio pode ser escrita, usando Eq.(A.8) do Apêndice A, como[3]

$$dS^\mu/d\tau = - \Gamma^\mu_{\alpha\beta} S^\alpha dx^\beta/d\tau \qquad (3.1).$$

Para calcular $dS^\mu/d\tau$ vamos considerar um vetor unitário **n** = **S**/S que coincide com a direção e sentido do eixo de spin do giroscópio. Assim, ao invés de (3.1) teremos[3]

$$dn^\mu/d\tau = - \Gamma^\mu_{\alpha\beta} n^\alpha dx^\beta/d\tau \qquad (3.2).$$

Os vetores $n^\alpha$ e $x^\alpha$ estão referidos no CRS. O vetor $n^\alpha$ que dá a direção do spin no espaço não tem uma parte temporal, ou seja, $n^4 = n^o = 0$ e, além disso, $dx^\beta/d\tau = (1,0,0,0)$. Nessas condições (3.2) fica dada por

$$dn^k/d\tau = - \Gamma^k_{po} n^p \qquad (k,p=1,2,3) \qquad (3.3).$$

Na Figura 2 vemos[3] o giroscópio, num dado instante, representado por um ponto G na origem do sistema de coordenadas $x^\mu$, descrevendo uma geodésica circular em volta da Terra. As coordenadas $x^\mu$ definem o CRS instantâneo de referência no qual o giroscópio está em repouso no ponto G em um dado instante. A velocidade instantânea v do ponto G, que está ao longo de x, em relação ao sistema $X^\mu$ fixo na Terra é

$$v = (GM/R)^{1/2} \qquad (3.4),$$

onde $R = R_o$ é distância do giroscópio ao centro da Terra. A aceleração radial a de G, que na Fig.2 está ao longo de z, é dada por

$$a = -GM/R^2 \qquad (3.5).$$

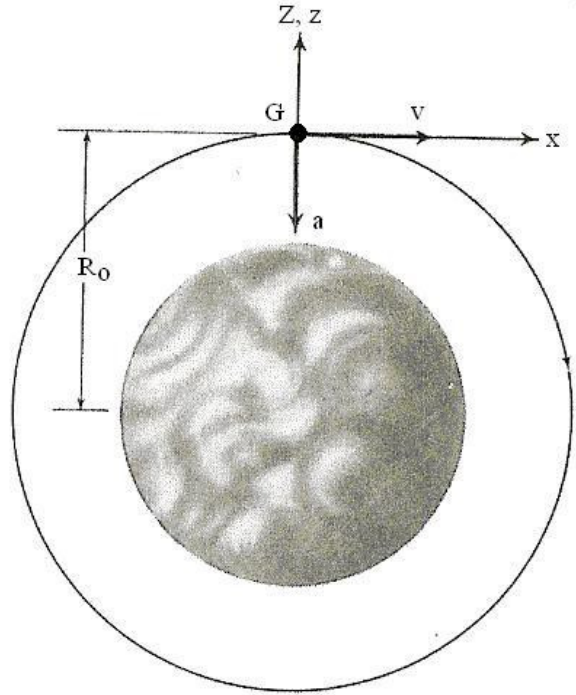

**Figura 2.** O giroscópio G se deslocando em uma geodésica circular em torno da Terra; G está na origem de um *referencial em queda (RQL)* (x,y,z) ou *comoving referential system*(CRS).

Usando as transformações de Lorentz[3,5] de $x^\mu$ para $X^\mu$, num instante inicial, desprezando a aceleração do sistema $x^\mu$ e levando em conta que $v/c \ll 1$ obtemos

$$X^o = x^o + (v/c)\, x^1, \qquad (3.6)$$

$$X^1 = x^1 + (v/c)\, x^o \qquad (3,7)$$

$$X^2 = x^2 \qquad (3.8)$$

$$X^3 = x^3 + R \qquad (3.9).$$

Como queremos calcular as derivadas (3.3) nós precisamos também definir o CRS num instante seguinte. Obviamente, para evitar uma rotação espúria de $n^\alpha$ o CRS no instante seguinte deve ter a mesma orientação do CRS no instante anterior. Mas, como sabemos da Relatividade Restrita mesmo se sucessivos CRS não tenham rodado um em relação ao outro, eles, entretanto, rodaram em relação ao referencial $X^\mu$ (vide Apêndice B).

Num pequeno intervalo de tempo $x^o$ o ângulo de rotação $\Delta\theta$ entre $X^\mu$ e $x^\mu$ é dado por $\Delta\theta = vax^o/2c^2$, num sentido anti−horário na Fig.2. Esta rotação gera o que chamamos de *precessão de Thomas* (Apêndice B) que aparece devido a uma transformação de Lorentz de ângulos[22] entre $X^\mu$ e $x^\mu$. Como nós pretendemos calcular a precessão do giroscópio no referencial que não roda relativamente a $x^\mu$, é preciso subtrair o efeito da precessão de Thomas do CRS. Assim, as equações (3.5)– (3.8) ficam escritas como

$$X^1 = x^1 + (v/c) x^o - vax^o x^3/2c^2 \qquad (3.10)$$

$$X^3 = x^3 + R + vax^o x^1/2c^2 \qquad (3.11).$$

Os últimos termos de (3.10) e (3.11) representam as rotações espaciais ordinárias devido a $\Delta\theta = vax^o/2c^2$, em primeira ordem em $\Delta\theta$.

Agora iremos usar (3.6),(3.8),(3.10) e (3.11) para calcular os $g_{\mu\nu}$ nas coordenadas $x^\mu$ levando em conta os $G_{\mu\nu}$ nas coordenadas $X^\mu$ que no limite de campos fracos da Terra, $2GM/c^2r \ll 1$, são definidos pelo invariante $ds^2$ dado, de acordo com (I.2), por $ds^2 \approx (1- 2F/R) c^2 dt^2 - (1 + 2F/R)(dX^2 +dY^2+dZ^2)$, onde fizemos $F = GM/c^2$ sendo $R = (X^2+Y^2+Z^2)^{1/2}$ Ou seja, $G_{44}=(1- 2F/R)$ e $G_{11}= (1 + 2F/R)$.

Os $g_{\mu\nu}$ no CRS são calculados através da equação de transformação[1d] dada por $g_{\mu\nu} = (\partial X^\alpha/\partial x^\mu)(\partial X^\beta/\partial x^\nu) G_{\alpha\beta}$. Com um cálculo simples, mas, relativamente extenso obtemos (usando o Maple 9), desprezando termos da ordem aF e $a^2$ e pondo $\nu = v/c$,

$g_{11}= -1- 2F/R+ \nu^2$     $g_{12}= 0$     $g_{13}= 0$     $g_{14}= -4F\nu/R + \nu ax^3/2$

$g_{21}= 0$     $g_{22}= -1-2F/R$     $g_{23}= 0$     $g_{24} = 0$

(3.12),

$g_{31}= 0$     $g_{32}= 0$     $g_{33}=-1-2F/R$     $g_{34}= -\nu ax^1/2$

$g_{41}= \nu ax^3/2 -4F\nu/R$     $g_{42}= 0$     $g_{43}=-\nu ax^1/2$     $g_{44}= 1- 2F/R- \nu^2$

Para calcular $dn^k/d\tau = -\Gamma^k_{po} n^p$, conforme (3.2), temos ainda de obter os símbolos de Christoffel[1]

$$\Gamma^\alpha_{\mu\nu} = \{^\alpha_{\mu\ \nu}\} = (g^{\alpha\lambda}/2)(g_{\lambda\mu,\nu} + g_{\lambda\nu,\mu} - g_{\mu\nu,\lambda}) \qquad (3.13)$$

usando os $g_{\mu\nu}$ dados em (3.12) lembrando[1] que $g^{\mu\nu} = M_{\mu\nu}/|g|$ onde g é o determinante de $g_{\mu\nu}$ e $M_{\mu\nu}$ é o determinante menor de $g_{\mu\nu}$ em g. Conforme (3.3) só temos de calcular os coeficientes $\Gamma^k_{po}$ onde k e p = 1,2 e 3. Como os $g_{\mu\nu}$ não dependem de $x^4$, ou seja, $g_{\lambda\mu,4} = 0$ a (3.13) fica escrita, levando em conta que $g_{ij} = g_{ji}$ :

$$\Gamma^k{}_{p4} = \Gamma^k{}_{po} = (g^{k\lambda}/2)( g_{\lambda 4, p} - g_{p4, \lambda} ) = (g^{k\lambda}/2)( g_{4\lambda, p} - g_{4p, \lambda} ) \quad (3.14),$$

De (3.4), como $n^o = 0$ temos $\Gamma^k{}_{44} = \Gamma^k{}_{oo} = 0$. Considerando esta última equação pode-se mostrar que[24]

$$\Gamma^k{}_{p4} = \Gamma^k{}_{po} = ( g_{4k, p} - g_{4p,k} )/2 \quad (3.15).$$

Como k, p = 1,2 e 3 verificamos usando os $g_{\mu\nu}$ g dados por (3.12) e (3.15) que $\Gamma^1{}_{34}$ e $\Gamma^3{}_{14}$ são os únicos não nulos. Como $F = GM/c^2$ teremos $F/R= \Phi(R)/c^2$, onde $\Phi(R) = GM/R$ é o potencial gravitacional. Assim, $g_{41}$ fica escrito na forma $g_{41}= vax^3/2 - 4\Phi v/c^2$. Os coeficientes da transformação $x^\mu$ para $X^\mu$ foram calculados levando em conta somente termos lineares $x^\mu$ nas vizinhanças da origem do CRS onde está colocado o giroscópio. Como no cálculo dos símbolos de Christoffel $\Gamma^1{}_{34}$ e $\Gamma^3{}_{14}$ derivamos $g_{\alpha\beta}$ em primeira ordem de $x^\alpha$ torna-se necessário expandir $\Phi(R)$ nas vizinhanças da origem do CRS em primeira ordem de $x^\alpha$. Assim, poremos $\Phi(R + \delta) \approx \Phi(R) + (\partial\Phi/\partial R) \delta$, onde $\delta \sim x^\alpha$. Desse modo vemos que

$$\Gamma^3{}_{14} = (g_{43, 1} - g_{41,3} )/2 = -va/2 - 2vGM/R^2c^2 = -(3/2)vGM/R^2c^2 \quad (3.16),$$

levando em conta que a aceleração centrípeta a do giroscópio é dada por $a = -GM/R^2$. Como $\Gamma^3{}_{14} = - \Gamma^1{}_{34}$ as equações (3.3) ficam escritas como

$$dn^1/d\tau =(3/2)v(GM/c^3R^2)n^3 ,$$

$$dn^2/d\tau = 0 \quad\quad\quad e \quad\quad\quad (3.17)$$

$$dn^3/d\tau = -(3/2)v(GM/c^3R^2)n^1.$$

Na notação vetorial 3-dim temos: **S** = S **n**, a aceleração gravitacional **A** = grad($\Phi$) =A (**R**/R), **R** a posição do giroscópio em relação ao centro da Terra e **v** a velocidade tangencial do giroscópio ao longo de sua trajetória circular. Assim, a representação vetorial 3-dim de (3.17) é dada por

$$d\mathbf{S}/d\tau = \mathbf{\Omega}_G \times \mathbf{S} \quad (3.18),$$

onde a "velocidade angular geodética" $\mathbf{\Omega}_G$ é definida por

$$\mathbf{\Omega}_G = (3/2c^3) \mathbf{A} \times \mathbf{v} \quad (3.19).$$

Em termos do momento angular orbital do giroscópio **L = R** x m**v** a (3.18) fica escrita como

$$d\mathbf{S}/d\tau = (3/2) (GM/mR^3c^3) \mathbf{L} \times \mathbf{S} \quad (3.20),$$

onde agora $\boldsymbol{\Omega}_G$ é dada por

$$\boldsymbol{\Omega}_G = (3/2)(GM/mR^3c^3)\,\mathbf{L} \qquad (3.21),$$

mostrando claramente a precessão do spin **S** do giroscópio em torno do momento angular **L** do satélite em torno da Terra. Notemos que a precessão geodética não depende do "spin" da Terra ($\mathbf{J} = \mathbf{S}_e$) como ocorre no efeito L–T.

Na Figura 3 mostramos[3] dois giroscópios em uma órbita polar a uma altura de ~8 000 m acima da superfície terrestre. No caso do primeiro giroscópio quando **S** está no plano da órbita verifica-se, usando a (3.21), que $\Omega_G = (3/2)(GM/R^2c^3)(GM/R)^{1/2} \sim 6.9"$/ano. Para o segundo giroscópio da figura com $\mathbf{S} \parallel \mathbf{L}$ temos $\Omega_G = 0$. A precessão de $0.05"$/ano vista na Fig.3 para esse segundo giroscópio corresponde a uma precessão do spin **S** do giroscópio em torno do "spin" **J** da Terra, devida ao efeito L–T descrita por $\boldsymbol{\Omega}_{LT} = (G/c^2R^3)\,[\mathbf{J} - 3\boldsymbol{r}(\mathbf{J}\cdot\boldsymbol{r})]$, conforme (1.6).

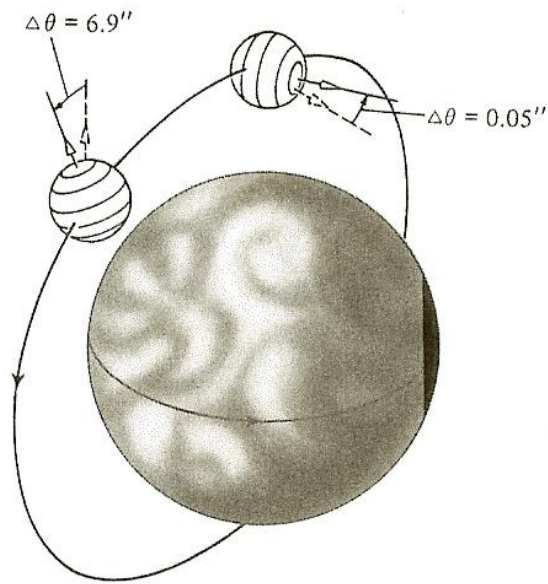

**Figura 3.** Dois giroscópios em órbita polar em torno da Terra. O primeiro que está com o spin **S** no plano da órbita, portanto, perpendicular a **L**, tem $\Omega_G = 6.9"$/ano. O segundo que está com $\mathbf{S}\parallel\mathbf{L}$, tem $\Omega_G = 0$ e $\Omega_{LT} = 0.05"$/ano.

É muito importante notarmos que devido à interação spin–órbita[23] que gera o EG que é dada por $U_{LS} = (3/2)(GM/mc^2r^3)\mathbf{L}\cdot\mathbf{S}$ a partícula não sofreria somente uma precessão de seu spin. Devido a $U_{LS}$ apareceria uma força (extremamente pequena) que provocaria um desvio de sua trajetória. Isto significa que uma partícula com spin num campo gravitacional sofreria um desvio de sua trajetória geodésica. Desse modo, partículas com spin (elétrons, prótons, nêutrons, etc...) num campo gravitacional não se moveriam exatamente ao longo de uma geodésica. Assim, o Princípio de Equivalência de Galileu não valeria para partículas com spin.

### 3.a) Comparação com resultados experimentais.Conclusões.

No caso geral, conforme é visto[25] na Figura 4, o spin **S** do giroscópio esférico irá precessionar em torno de seu momento angular orbital **L** devido à interação gravitacional entre **S** e **L** e, simultaneamente, em torno do eixo de rotação da Terra devido à interação entre **S** e o spin **J** = I$\omega$ da Terra. Enfim, a velocidade angular de precessão $\mathbf{\Omega} = \mathbf{\Omega}_G + \mathbf{\Omega}_{LT}$ resultante desses dois efeitos é dada por (3.21) e (1.11), respectivamente,

$$\mathbf{\Omega} = \mathbf{\Omega}_G + \mathbf{\Omega}_{LT} = (3/2)(GM/mR^3c^3)\,\mathbf{L} + (GI/c^2R^3)\,[\mathbf{\omega} - 3r(\mathbf{\omega}\cdot r)] \quad (3.22).$$

Na Figura 4 vemos um esboço de como estariam colocados os giroscópios esféricos criogênicos de grande precisão[25] se deslocando ao longo de uma baixa órbita polar (~642 km) em torno da Terra e uma estrela guia (IM Pegasi) como um referencial inercial. O rotor do giroscópio é de quartzo homogêneo com um diâmetro ~4 cm. Um desvio da forma esférica maior do que $10^{-6}$ cm é inaceitável. O efeito Geodético teria um desvio angular anual ~ 6.6" e o efeito L−T ("Frame−dragging Effect") teria um desvio anual ~ 0.039". As medidas preliminares que foram feitas[25] mostram um bom acordo com as previsões teóricas obtidas com a (3.22). Apesar desses resultados serem muito promissores uma análise mais cuidadosa dos erros nas medidas está ainda em curso para se ter uma exata confirmação dos resultados.

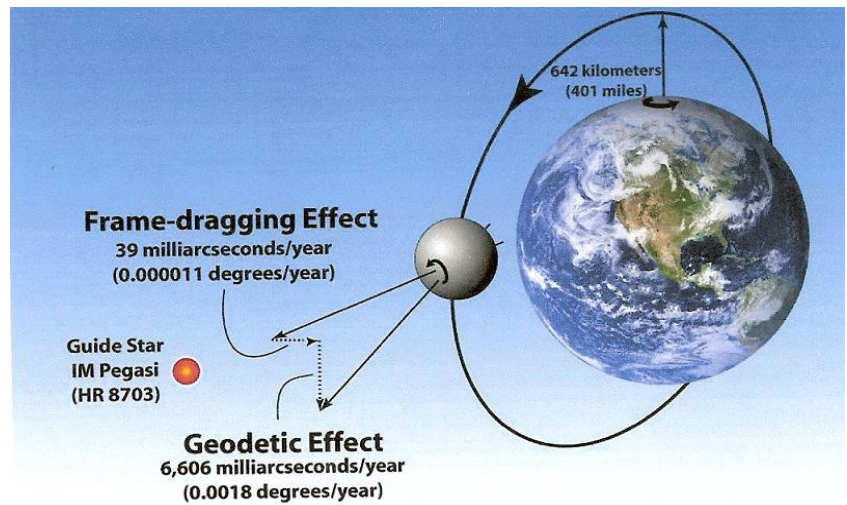

**Figura 4**. Mostra[25] o spin **S** do satélite em órbita polar sendo desviado simultaneamente pelo Efeito Geodético e pelo Efeito L−T ou "Frame −dragging Effect".

## Apêndice A. Coordenadas Geodésicas.

Conforme analisamos em um artigo anterior[1c] é sempre possível[3] encontrar coordenadas tais que em um dado ponto P a métrica do espaço–tempo é plana. Ou seja, dadas as coordenadas $x^\mu$ com uma métrica $g_{\mu\nu}(x)$ podemos sempre encontrar uma transformação linear para novas coordenadas $x'^\mu = b^\mu_\nu x^\nu$, onde $b^\mu_\nu$ são constantes, de tal modo que num certo ponto P tenhamos $g'_{\mu\nu}(P) = \eta_{\mu\nu}$, onde $\eta_{\mu\nu} = (-1,1,1,1)$ característico de um espaço–plano de Minkowski. Nas vizinhanças do ponto P onde o espaço é localmente plano temos um *referencial inercial local*. A transformação de coordenadas que satisfaz essas exigências é dada por[3]

$$x'^\mu = x^\mu - x^\mu(P) + (1/2) \Gamma^\mu_{\alpha\beta}(P) [x^\alpha - x^\alpha(P)][x^\beta - x^\beta(P)] \quad (A.1),$$

onde $x^\mu(P)$ são as antigas coordenadas do ponto P.

As coordenadas $x'^\mu$ são denominadas de *coordenadas geodésicas* e o referencial é denominado de *referencial geodésico* (RG). No ponto P as derivadas primeiras ordem de $g'_{\mu\nu}(x')$ e os símbolos $\Gamma^\mu_{\alpha\beta}(P)$ de Christoffel se anulam. Levando isso em conta na equação de uma geodésica (1.2), no referencial geodésico $x'^\mu$ obtemos

$$d^2 x'^\mu / ds^2 = 0 \quad (A.2).$$

Ou seja, uma partícula no ponto P se move com velocidade constante ou permanece em repouso em relação ao *referencial geodésico*. Isto significa que o referencial está instantaneamente em *queda livre* com a mesma aceleração que a partícula. Devemos enfatizar[3] que as coordenadas são geodésicas somente para um dado instante e um dado ponto. As derivadas da métrica são zero somente em certo ponto P do espaço–tempo. Se quisermos coordenadas geodésicas em um outro ponto diferente de P, será necessário realizar uma outra transformação de coordenadas $x'^\mu = b^\mu_\nu x^\nu$ para esse ponto.

Geometricamente a introdução das coordenadas geodésicas no ponto P corresponde a substituir o espaço curvo nesse ponto por um espaço plano tangente a P. As derivadas primeiras de $g'_{\mu\nu}(x')$ são nulas, mas, as de segunda ordem não são nulas.[3]

O ponto P é a origem do sistema de coordenadas geodésicas denominado *referencial geodésico* (RG) ou *referencial em queda livre* (RQL) ou, ainda, *comoving referential system* (CRS). Os referenciais em queda livre estão sempre ao longo de uma geodésica.

*Transporte paralelo ao longo de uma geodésica.*

Consideremos um 4−vetor contravariante constante $a'^\nu$ em um determinado ponto P de uma geodésica, que é a origem $x'^\mu = 0$ de um referencial geodésico. Façamos um *transporte paralelo*[1d,3] desse 4−vetor até um outro ponto P´distante $\delta x'^\mu$ de P da geodésica. O novo ponto P´sendo também a origem de um referencial geodésico. As componentes $a'^\nu$ do vetor não mudam com o transporte pois as coordenadas $x'^\mu$ são cartesianas. Ou seja, se o vetor original era $a'^\nu$ o novo vetor será $a'^\nu + \delta a'^\nu$, onde $\delta a'^\nu = 0$. Entretanto, as componentes $a^\nu$ do vetor em coordenadas curvilíneas mudam. As suas componentes antes e depois do transporte estão relacionadas pelas equações

$$a^\alpha = (\partial x^\alpha/\partial x'^\nu)_{x'=0} \, a'^\nu \qquad (A.3)$$

e

$$a^\alpha + \delta a^\alpha = (\partial x^\alpha/\partial x'^\nu)_{\delta x'} (a'^\nu + \delta a'^\nu) \qquad (A.4).$$

Para calcular (A.2) nós precisamos das derivadas $\partial x^\alpha/\partial x'^\nu$. A maneira mais simples de fazer isso é verificar que para pequenos valores de $x'^\mu$ nós podemos aproximar a (A.1) por

$$x^\mu \approx x^\mu(P) + x'^\mu - (1/2) \Gamma^\mu_{\alpha\beta}(P) \, x'^\alpha x'^\beta \qquad (A.5),$$

de onde tiramos $\partial x^\alpha/\partial x'^\nu \approx \delta^\alpha_\nu - \Gamma^\alpha_{\nu\beta}(P) \, x'^\beta$ que colocadas em (A.3) e (A.4) dão

$$a^\alpha = a'^\alpha \qquad (A.6)$$

e

$$a^\alpha + \delta a^\alpha = (a'^\alpha + \delta a'^\alpha) - \Gamma^\alpha_{\beta\nu}(P)(a'^\nu + \delta a'^\nu) \, \delta x'^\beta \qquad (A.7).$$

A diferença entre (A.6) e (A.7) dá

$$\delta a^\alpha = - \Gamma^\alpha_{\beta\nu}(P) \, a^\nu \, \delta x'^\beta \qquad (A.8).$$

De modo análogo, para um 4−vetor covariante $a_\alpha$ obtemos[3]

$$\delta a_\alpha = \Gamma^\nu_{\alpha\beta} \, a_\nu \, \delta x^\beta \qquad (A.9).$$

## Apêndice B. Precessão de Thomas.

Consideremos uma partícula P descrevendo uma órbita circular[23] em torno de um ponto O que é a origem de um referencial inercial (x,y) conforme vemos na Figura (B.1). Suponhamos que P esteja em repouso momentaneamente em relação aos referenciais $(x_1,y_1), (x_2,y_2)$ e $(x_3,y_3)$ nos instantes sucessivos $t_1 < t_2 < t_3$, respectivamente. Os instantes de tempo $t_i$ diferindo muito pouco uns dos outros. Os eixos ($x_i$, $y_i$) foram desenhados paralelos a (x,y). Entretanto, mostraremos que o observador em (x,y) vê os eixos ($x_2$,$y_2$) ligeiramente rodados em relação a (x,y) e os eixos ($x_3$,$y_3$) um pouco mais rodados ainda em relação a (x,y). Assim, ele vê que os eixos nos quais P se encontra instantaneamente em repouso estão precessionando em relação a O − embora os observadores instantaneamente em repouso em relação a P sustentem que cada referencial ($x_{i+1}$,$y_{i+1}$) esteja paralelo ao anterior ($x_i$,$y_i$).

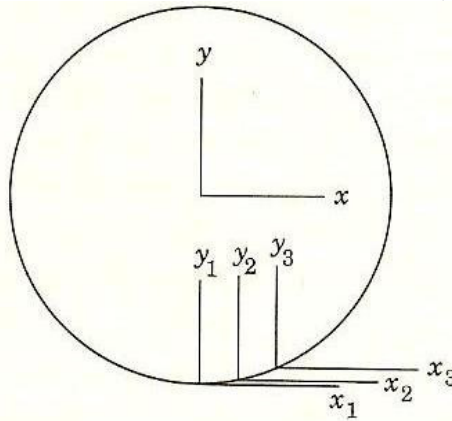

**Figura (B.1)**. O sistema inercial (x,y) e os "comoving referential systems" (CRS) ($x_i$,$y_i$)

    Na Figura B.2 vemos xy, $x_1y_1$ e $x_2y_2$ do ponto de vista do observador em $x_1y_1$. Como P se move com velocidade **v** em relação a O, os eixos (x,y) estão se movendo com velocidade −**v** em relação a ($x_1$,$y_1$). Visto de ($x_1$,$y_1$)

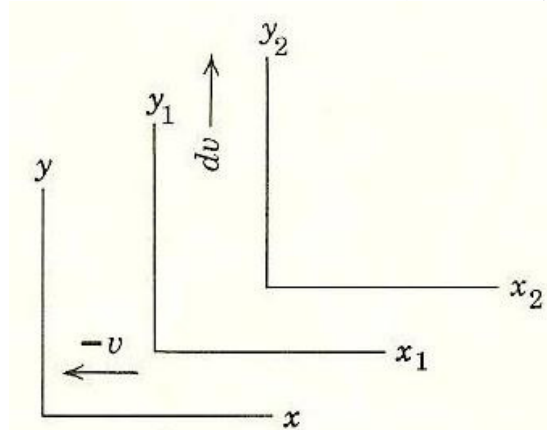

**Figura (B.2).** Os referenciais usados para calcular a precessão de Thomas, vistos do referencial ($x_1$,$y_1$).

o ponto P tem uma aceleração **a** em direção a O no sentido positivo de $y_1$. Se o intervalo de tempo $t_2 - t_1$ é muito pequeno, a mudança na velocidade de P nesse intervalo é dada por d**v** = **a** ($t_2 - t_1$) = **a** dt que será a velocidade de ($x_2$,$y_2$) em relação a ($x_1$,$y_1$). Usando as transformações de Lorentz[5,23] para velocidades, as componentes da velocidade $\mathbf{V}_a$ de $(x_2,y_2)$ observadas de $(x,y)$ são dadas por

$V_{ax} = (dv_x - v_x)/(1 - v_x dv_x/c^2) = (0 + v)/(1 - (-v).0/c^2) = v$

$V_{ay} = dv_y [1 - (v_x/c)^2]^{1/2}/(1 - v_x dv_x/c^2) = dv [1 - (v/c)^2]^{1/2}$ (B.1)

Com as transformações inversas, as componentes da velocidade $\mathbf{V}_b$ de $(x,y)$ vistas de $(x_2,y_2)$ são dadas por

$V_{bx} = v_x [1 - (dv_y/c)^2]^{1/2}/(1 - v_y dv_y/c^2) =$

$\quad -v [1 - (dv/c)^2]^{1/2}/(1 - 0.dv/c^2) = -v [1 - (dv/c)^2]^{1/2}$ (B.2).

$V_{by} = (v_y - dv_y)/(1 - v_y dv_y/c^2) = (0 - dv)/(1 - 0.dv_y/c^2) = -dv$

Analisando (B.1) e (B.2), conforme Figura (B.3), verificamos que $|\mathbf{V}_a| = |\mathbf{V}_b|$ de acordo com o postulado da equivalência dos referenciais inerciais. O ângulo $\theta_a$ entre $\mathbf{V}_a$ e o eixo x do referencial (x,y) é dado por

$\theta_a \approx \mathrm{tg}\, \theta_a = V_{ay}/V_{ax} = dv [1 - (v/c)^2]^{1/2}/v$

O ângulo $\theta_b$ entre o vetor $\mathbf{V}_b$ e o eixo $x_2$ do referencial $(x_2,y_2)$ é dado por

$\theta_b \approx \mathrm{tg}\, \theta_b = V_{by}/V_{bx} = dv/[v (1 - (dv/c)^2)^{1/2}] \approx dv/v$

Na Figura (B.3) vemos os referenciais $(x_2,y_2)$ e $(x,y)$ do ponto de vista de $(x,y)$.

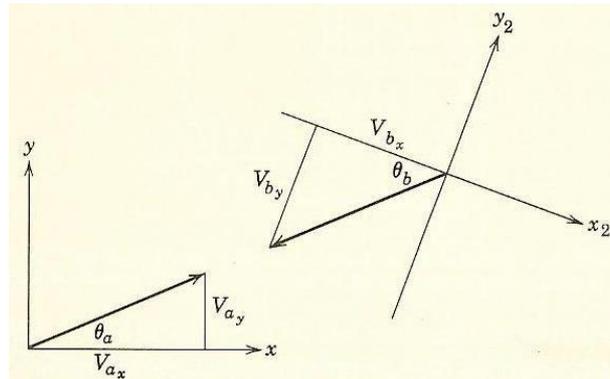

**Figura (B.3)**. Ilustração exagerada da precessão de Thomas.

De acordo com o postulado de equivalência $\mathbf{V}_a$ e $\mathbf{V}_b$ devem ter exatamente direções opostas, isto é, $\mathbf{V}_a = -\mathbf{V}_b$. Como os ângulos entre os eixos x e os vetores $\mathbf{V}$ não são os mesmos o referencial $(x_2,y_2)$ rodou em relação ao referencial $(x,y)$. O ângulo de rotação $d\theta$ é dado por

$d\theta = \theta_b - \theta_a = (dv/v) \{1 - [1 - (v/c)^2]^{1/2}\}$ ,

que no limite (v/c) << 1 dá
$$d\theta \approx dv\, v^2/2vc^2 = vdv/2c^2 = va\, dt/2c^2,$$

levando em conta que dv = a dt. Desse modo vemos que os eixos em relação aos quais o ponto P está em repouso precessiona, em relação a O (origem do referencial inercial (x,y)) com uma velocidade angular

$$\omega_T = d\theta/dt = va/2c^2 \qquad (B.3),$$

denominada de velocidade angular de *precessão de Thomas*. Através das figuras (B.1)–(B.3) vemos que ela é dada vetorialmente por

$$\boldsymbol{\omega}_T = -(1/2c^2)\, \mathbf{v} \times \mathbf{a} \qquad (B.4).$$

## REFERÊNCIAS

[1] M.Cattani. (1a) http://arxiv.org/abs/1005.4314/ (2010).
(1b) http://arxiv.org/abs/1007.0140/ (2010)
(1c) http://arxiv.org/abs/1010.0241/ (2010); (1d) RBEF 20, 27 (1998).
[2] C.W.Misner, K.S.Thorne and J.A.Wheeler."Gravitation", Freeman (1970).
[3] H.C.Ohanian. "Gravitation and Spacetime".W.W.Norton (1976).
[4] I.R.Kenyon. "General Relativity", Oxford University Press (1990).
[5] L.Landau et E.Lifchitz."Théorie du Champ", pág.421. Éditions de la Paix (1964).
[6] J.Lense and H.Thirring.Phys.Zeits.19, 156 (1918).
[7] R.P.Kerr. Phys.Rev.Lett.11,237(1963).
[8] R.H.Boyer and R.W.Lindquist.J.Math.Phys.8,265(1967).
[9] M.Rees, R.Ruffini and J.A. Wheeler. "Black Holes, Gravitational Waves and Cosmology". Gordon and Breach, NY(1974).
[10] W.Rindler. "Relativity, Special, General and Cosmological", Oxford (2008).
[11] H.Yilmaz. "Theory of Relativity and the Principles of Modern Physics", Blaisdell Publishing Company, NY(1965).
[12] H.Goldstein."Classical Mechanics", Addison–Wesley (1959).
[13] L.Landau e E.Lifchitz."Mecânica". Hemus–Livraria e Editora (SP) (1970).
[14] I.Ciufolini. Phys.Rev.Lett.56, 278(1987).
[15] B.Davies. Am.J.Phys.51,909(1983).
[16] K.A.Milton. Am.J.Phys.42, 911(1974).
[17] J.Schwinger."Particles, Sources and Fields" vol.I,sec.2–4. Addison–Wesley(1971).
[18] I.Ciufolini. Class.Quantum Grav.2369(2000).
[19] G.E.Pugh.WSEG Research Memo11, U.S. Dep. of Defense (1959).
[20] L.I.Schiff. Proc.Nat.Acad.Sci.46, 871(1960); Phys.Rev.Lett.4, 215(1960).
[21] **W.** de Sitter. Mon. Not. Roy. Astron. Soc. **77**: 155 (1916).
[22] Vide Ref.3, pag.292–293.
[23] R.M.Eisberg. "Fundamentals of Modern Physics", John Wiley &Sons (1964).
[24] N.Ashby. "Gyroscope Precessions and Gravitometric Effects" (2002) http://www.colorado.edu/physics/phys7840/phys7840_fa02/precess/precess.html
[25] J.W.Conklin. "Sixth International Conference on Gravitation and Cosmology IOP Publishing". Journal of Physics: Conference Series **140** (2008) 012001.